\documentclass[iop, numberedappendix]{mn2e}

\usepackage[pdftex]{graphicx}

\usepackage{amsmath, amsthm, amssymb}
\let\bibhang\relax
\usepackage{natbib}
\usepackage{subfigure}
\usepackage{upgreek}
\usepackage{accents}
\usepackage{color}
\usepackage{hyperref}
\usepackage{dblfloatfix}
\usepackage[T1]{fontenc}
\usepackage{aecompl}
\usepackage{enumitem}
\usepackage[normalem]{ulem}

\hypersetup{
  colorlinks   = true, %Colours links instead of ugly boxes
  urlcolor     = blue, %Colour for external hyperlinks
  linkcolor    = blue, %Colour of internal links
  citecolor   = blue %Colour of citations
}

\newcommand{\AUV}{A_{\text{UV}}}

\newcommand{\microns}{\mu \text{m}}

\newcommand{\lambdaObs}{\lambda_{\text{o}}}
\newcommand{\lambdaEm}{\lambda_{\text{e}}}

\newcommand{\angstrom}{\text{\normalfont\AA}}

\newcommand{\mAB}{m_{\text{AB}}}

\newcommand{\zpr}{z^{\prime}}

\newcommand{\obs}{\text{obs}}

\newcommand{\zobs}{z_{\obs}}

\newcommand{\HI}{\text{H} {\textsc{i}}}

\newcommand{\Lya}{\text{Ly-}\alpha}

\newcommand{\MUV}{M_{\text{UV}}}

\newcommand{\mH}{m_{\text{H}}}

\newcommand{\zrei}{z_{\text{rei}}}
\newcommand{\zgal}{z_{\text{gal}}}

\newcommand{\fcoll}{f_{\text{coll}}}

\newcommand{\Mmin}{M_{\min}}

\newcommand{\fesc}{f_{\text{esc}}}

\newcommand{\Msun}{M_{\odot}}

\newcommand{\zform}{z_{\text{form}}}

\title[EoR Feedback in NIRB]{Signatures of reionization feedback in the near-infrared background}
\author[Mirocha, Liu, \& La Plante]{
Jordan Mirocha,$^{1}$\textsuperscript{\thanks{jordan.mirocha@mcgill.ca}}
Adrian Liu$^{1}$, and Paul La Plante$^{2}$\textsuperscript{\thanks{BCCP Fellow}} \\
$^{1}$McGill University Department of Physics \& McGill Space Institute, 3600 Rue University, Montr\'eal, QC, H3A 2T8 \\
$^{2}$Department of Astronomy and Berkeley Center for Cosmological Physics, University of California Berkeley, Berkeley, CA 94720, USA \\
}

\begin{document}

\pagerange{\pageref{firstpage}--\pageref{lastpage}} \pubyear{2020}
\maketitle

%%%
%% Abstract
%%%
\begin{abstract}
The reionization of the intergalactic medium at redshifts $z\gtrsim 6$ is expected to have a lasting impact on galaxies residing in low-mass dark matter halos. Unable to accrete or retain gas photo-heated to temperatures $T \gtrsim 10^4$ K, the star formation histories of faint galaxies in the early Universe are expected to decline as they exhaust their gas supply, resulting in a turn-over in the galaxy luminosity function (LF) and a potential solution to the missing satellites problem in the local group. Unfortunately, there are several challenges to constraining `reionization feedback' empirically, most notably that galaxies in low-mass halos are intrinsically faint, and that there are other physical mechanisms capable of inducing a turn-over in the LF. In this work, we investigate a new signature of reionization feedback that is in principle distinct from other processes: as faint galaxies are quenched by reionization, their stellar populations passively age and grow redder while the brighter galaxies nearby continue to form stars at an increasing rate and so remain relatively blue. We find that this contrast, between quenched and un-quenched galaxies induces a scale and colour-dependent signature in the present-day near-infrared background comparable to the expected sensitivity of NASA's upcoming SPHEREx mission. Whereas models with pure mass suppression largely affect the signal at wavelengths $\lesssim 2 \microns$, $\sim 5$\%-level differences in the background persist out to $\simeq 5 \microns$ for reionization feedback models. Finally, the power spectra of intensity \emph{ratio} maps exhibit larger variations, and may thus be a promising target for future analyses.
\end{abstract}

%%%
%% Introduction
%%%
\section{Introduction} \label{sec:intro}
The Epoch of Reionization (EoR), when ultraviolet (UV) photons from the first galaxies ionized the intergalactic medium (IGM), is an important milestone in the cosmic timeline. Not only does the transformation from neutral to ionized open up new windows into galaxy formation via, e.g., $\Lya$ emission, 21-cm tomography, and anisotropies in the cosmic microwave background (CMB), it is also expected to have a lasting impact on small structures throughout the Universe. As the ultraviolet background (UVB) from early sources heats and ionizes the IGM, sufficiently shallow dark matter (DM) potential wells hosting low-mass galaxies will struggle to accrete new gas, and may even lose what gas they have to photoevaporation. Dwindling gas supplies will reduce star formation rates, with the fraction of galaxies quenched by reionization presumably rising continually as the volume fraction of ionized gas approaches unity at $z \sim 5-6$.

Given that reionization heats the IGM to $\sim 10^4$ K, one might expect that only halos with virial temperatures less than $\sim 10^4$ K fall victim to reionization feedback. However, in reality the situation is more complicated, giving rise to a substantial literature focused on the critical scale of UVB-driven suppression of galaxy formation, which goes back several decades \citep{Couchman1986,Rees1986,Ikeuchi1986,Efstathiou1992,Shapiro1994,Quinn1996,Thoul1996,Barkana1999}, and remains an active area of research \citep[e.g.,][]{Benson2002a,Susa2004,Okamoto2008,Simpson2013,Noh2014,Onorbe2015,Dawoodbhoy2018,Katz2020}. Beyond the virial temperature argument, the Jeans mass is a natural next scale to consider. However, a key insight is that, while the critical suppression scale is \textit{related to} the Jeans mass, it is not \textit{equal to} the Jeans mass \citep{GnedinHui1998}. Because the baryons respond to temperature changes on a dynamical timescale, the more relevant `filtering scale` lags behind the Jeans scale, and the suppression occurs in $M_h \sim 10^8-10^9 M_{\odot}$ objects, about an order-of-magnitude smaller than naively expected from Jeans scale arguments \citep{Gnedin2000}.

Despite the growing theoretical understanding of UVB-induced feedback on galaxy formation, empirical evidence for reionization feedback remains scarce. Halos with masses of $\sim 10^8-10^9 M_{\odot}$ likely host galaxies with absolute magnitudes of $-12 \lesssim \MUV \lesssim -9$ in the rest-UV continuum, $\lambdaEm \sim 1600 \angstrom$ \citep[e.g.,][]{Behroozi2019,Yung2019}, or perhaps $\MUV \sim -15$ in extreme cases \citep{Yue2016a}. The canonical $\MUV \sim -12$ cutoff is beyond the reach of all but the most highly lensed fields used for high-$z$ galaxy surveys, results from which are still too uncertain to conclusively identify a turn-over in the galaxy luminosity function (LF) at high redshift \citep{Atek2018}. Given the challenge of detecting any LF turn-over at high redshifts directly, it seems prudent to explore other observable signatures of reionization feedback.

The local group provides perhaps the most powerful constraints on reionization feedback so far assembled. Models have long predicted an abundance of small halos but not necessarily an abundance of faint galaxies \citep[e.g.][]{White1978,Dekel1986,Kauffmann1993}. This implies a `missing satellites' problem for the Milky Way \citep{Klypin1999,Moore1999}, as the local group has relatively few dwarfs \citep{Tolstoy2009,McConnachie2012}. There are many potential solutions, e.g., warm dark matter \citep[e.g.,][]{Bode2001}. Alternatively, there may be no shortage of halos around the Milky Way, but rather a low fraction of halos that host luminous galaxies \citep[e.g.,][]{Nadler2020}. Baryonic processes, such as gas stripping and reionization feedback, can indeed reduce the number of satellites relative to the halo abundance and modulate the luminosity function, bringing models into closer agreement with observations \citep{Bullock2000,Somerville2002,Benson2002b,Wyithe2006,Brooks2013,Bose2018,Kravtsov2021}.  Reconstructions of the star formation histories of some nearby dwarfs do show evidence of a decline in star formation beginning at $z \sim 6$ \citep{Weisz2014SFHsII,Brown2014}, suggestive of reionization feedback, though there are also dwarfs that have undergone non-negligible star formation well past reionization and/or had their star formation quenched relatively recently \citep{HurleyKeller1998,Grebel2004,Cole2007,Young2007,Monelli2010,Cole2014,McQuinn2015,Weisz2015}.

While thirty-meter-class telescopes will expand the local volume in which such analyses can be performed \citep{WeiszBK2019}, more `global' constraints on reionization feedback are still worth pursuing. For example, the bulk properties of the low-mass galaxy population may be indirectly inferred via the timing, duration, and topology of reionization. Current constraints from CMB observations limits star formation in low-mass halos through their combined impact on the reionization history \citep[e.g.][]{Miranda2017,Wu2021}, while the decline of the ionizing background after reionization inferred from $\Lya$ forest measurements could indicate a commensurate decline in the emissivity of low-mass sources \citep[e.g.,][]{Ocvirk2021}. 21-cm measurements offer a conceptually similar, but in principle more precise probe of the low-mass galaxy population \citep[e.g.,][]{Qin2021,GesseyJones2022}. However, in each of these cases there are degeneracies between the suppression scale, star formation efficiency, and escape fraction of ionizing photons. Even if a turn-over in the galaxy LF can be inferred from 21-cm observations, it may be difficult to distinguish different feedback scenarios, at least on the scales accessible to current and near-future experiments \citep{Hutter2020}, essentially because star formation is already inefficient in low-mass galaxies \citep{Wyithe2013}.
Given the potential complications of using the local group and reionization to constrain the low-mass galaxy population, the near-infrared background (NIRB) may offer a powerful probe of reionization feedback, as it is sensitive largely to the \textit{non-ionizing} rest-ultraviolet and optical continuum of high-$z$ sources. Many studies have explored how the properties of early galaxies affect the NIRB, both its mean spectrum and fluctuations \citep[e.g.,][]{Santos2002,Cooray2004,Kashlinsky2007,Fernandez2006,Fernandez2010,Helgason2015,Cooray2012,Yue2013Galaxies,Yue2016b,Sun2021}. Though low-$z$ contamination remains a concern \citep{Zemcov2014,Feng2019}, fiducial models for Pop~II galaxies predict signals well above the nominal sensitivity of SPHEREx \citep{Dore2014,Dore2016}, while a detection of Pop~III contributions will require rather extreme scearios with very efficient star formation and exquisite component separation \citep{Sun2021}.

In this work, we focus specifically on the signature of reionization feedback, neglecting Pop~III stars and leaving the parameters of Pop~II sources fixed; we vary only the treatment of suppression. The assumption of a critical mass threshold below which star formation ceases is built-in to every reionization model, and is known to affect many observables because it limits the overall number of galaxies and modulates the typical bias of sources. Adding a \textit{position-dependent} suppression, as one would expect from reionization, does indeed modulate fluctuations in the NIRB as well, as pointed out by \citet{Fernandez2012}. Here, we focus on another aspect of the problem, which to our knowledge has been neglected in all previous work. Though to a good approximation the rest-ultraviolet emission of quenched galaxies will vanish immediately, their rest-optical emission will persist, and so not be well-modeled by a simple suppression scale. In other words, the degree of suppression is mass, position, and colour-dependent. Solving this problem requires forward modeling the spectra of all simulated objects in detail over their entire history, and thus adds to the already non-trivial computational cost associated with modeling galaxy assembly and reionization. However, this could be a timely consideration given that, e.g., SPHEREx's spectral coverage extends to $5 \mu\rm{m}$, and so includes the rest-optical emission of reionization era galaxies.

One might naively expect the effect of stellar aging to simply weaken NIRB fluctuations, since an otherwise dark patch of the Universe will now be luminous, and thus have less contrast between star-forming galaxies. While this is likely true to some extent, there is a new opportunity to unlock constraining power in colour space. Given the contrast betweeen a sea of passively-aging `quenched' galaxies and the massive galaxies unaffected by reionization will leave a characteristic scale- and colour-dependent signature in the present-day NIRB. A scale dependence is expected given the difference in clustering between bright and faint galaxies, as well as the characteristic scale associated with ionized bubbles during reionization, while a colour-dependence is expected given the difference in the spectra of quenched and unquenched objects as well as the timing of reionization. Though far from a `clean' probe, given the mixing of spectral and temporal information, the diffuse NIRB's sensitivity to galaxies \textit{and} reionization may allow it to uniquely identify the cause of any turn-over in the galaxy LF, as other potential sources of a turn-over (e.g., warm dark matter, stellar feedback) will not correlate with the reionization history.

We outline our approach to modeling high-$z$ galaxies, reionization, and the NIRB in Section \ref{sec:methods}. We present our main results in Section \ref{sec:results}, chiefly the expected signature of reionization feedback in the NIRB, and discuss the broader implications of our work and conclude in Section \ref{sec:conclusions}.

We adopt AB magnitudes throughout \citep{Oke1983}, i.e.,
\begin{equation}
    M_{\lambda} = -2.5 \log_{10} \left(\frac{f_{\lambda}}{3631 \ \mathrm{Jy}} \right)
\end{equation}
and use a \textit{Planck} 2015 cosmology \citep{Planck2015}.

%%%
%% Methods
%%%
\section{Methods} \label{sec:methods}
In this section we outline our pipeline for modeling the NIRB. We first describe the $N$-body simulations and galaxy modeling (\S\ref{sec:galaxies}), followed by our approach to reionization quenching and NIRB synthesis in \S\ref{sec:quenching} and  \S\ref{sec:nirb}, respectively. Readers familiar with the subject may skip to illustrative examples of model galaxies in Fig. \ref{fig:aging_seds} and \ref{fig:aging_uvlfs}, the suppression model in Fig. \ref{fig:mcrit}, and example NIRB mocks in Fig. \ref{fig:mock_nirb}.

%%
% N-body
\subsection{Model for Galaxies} \label{sec:galaxies}
The mock NIRB maps we present in this work are based on the semi-analytic model presented in \citet{Mirocha2021b}. We defer the details to that work, reviewing here only the most salient details.

Our model follows what is now a common semi-empirical approach, in which the star formation efficiency (SFE) of galaxies is left to vary freely in fits to the observed rest-ultraviolet luminosity function \citep[UVLF; e.g.,][]{Mason2015,Mashian2016,Sun2016,Mirocha2017,Tacchella2018,Behroozi2019}. We model the SFE as a double power-law in halo mass, $M_h$, and assume it is invariant with cosmic time in this work. We employ a simple -- but self-consistent -- model for dust reddening, and jointly fit UVLFs \citep[from][]{Bouwens2015} and UV colour-magnitude relations ($\MUV$-$\beta$) \citep[from][]{Bouwens2014}. As described in \citet{Mirocha2020a}, this results in systematically shallower relationships betweeen UV extinction and UV magnitude, $\AUV(\MUV)$, than those predicted from the \citet{Meurer1999} relation derived from $z \sim 3$ galaxy samples. The models agree well with measurements based on forward-modeling the SEDs of observed sources, in which $\beta$ is inferred within the \citet{Calzetti1994} spectral windows of the best-fitting SED \citep{Finkelstein2012}, rather than estimated from the broadband photometry, as in \citet{Bouwens2014}.

The key difference between previous work with this model, as implemented in \textsc{ares}\footnote{\url{https://github.com/mirochaj/ares}} \citep{Mirocha2017,Mirocha2020a}, and the current work is that, rather than generating a population of halo growth histories from the halo mass function and mean halo mean mass accretion (MAR) \citep[following][]{Furlanetto2017}, we take halo assembly histories directly from the merger tree constructed from an $N$-body simulation. In this work we draw results from a single $(80 \ \rm{Mpc} / h)^3$ box run with $2048^3$ particles with the P$^3$M algorithm described in \citet{Trac2015}. Halo catalogs are generated on-the-fly every 20 Myr in cosmic time, initially using a friend-of-friends algorithm \citep[FOF;][]{Davis1985}, and subsequently refined using a spherical overdensity approach. We construct merger trees using custom routines similar to other techniques documented in the literature \citep[e.g.,][]{Lacey1994,Somerville1999,Behroozi2013Trees,Poole2017}. The mass function of DM halos in the simulation agrees well with the \citet{Tinker2010} form, as generated with the \textsc{hmf}\footnote{\url{https://hmf.readthedocs.io/en/latest/}} code \citep{Murray2013}. The MAR of pristine material from the IGM, inferred via the change in halo masses between time-steps (minus the mass growth from mergers), agree well with past work \citep{McBride2009,Trac2015}, and dominate the mass growth of halos at all redshifts explored in this study, $z \gtrsim 5$  \citep[see Figs. 2-3 in][]{Mirocha2021b}.

In \citet{Mirocha2021b}, we showed that the default \textsc{ares} model \citep[run with the mean MAR of][]{Furlanetto2017} over-estimates the SFE because its idealized halos do not grow as rapidly as halos drawn from $N$-body simulations. As a result, galaxies hosted in simulated halos are brighter and bluer, holding all else constant. Fortunately, the SFE and dust scale length can be re-scaled slightly in order to bring the simulation-based model back into agreement with observed UVLFs and UV colours. We employ the same rescaling in this work for consistency, meaning our mock galaxies are identical to those in \citet{Mirocha2021b} before accounting for suppression.

\begin{figure*}
\begin{center}
\includegraphics[width=0.98\textwidth]{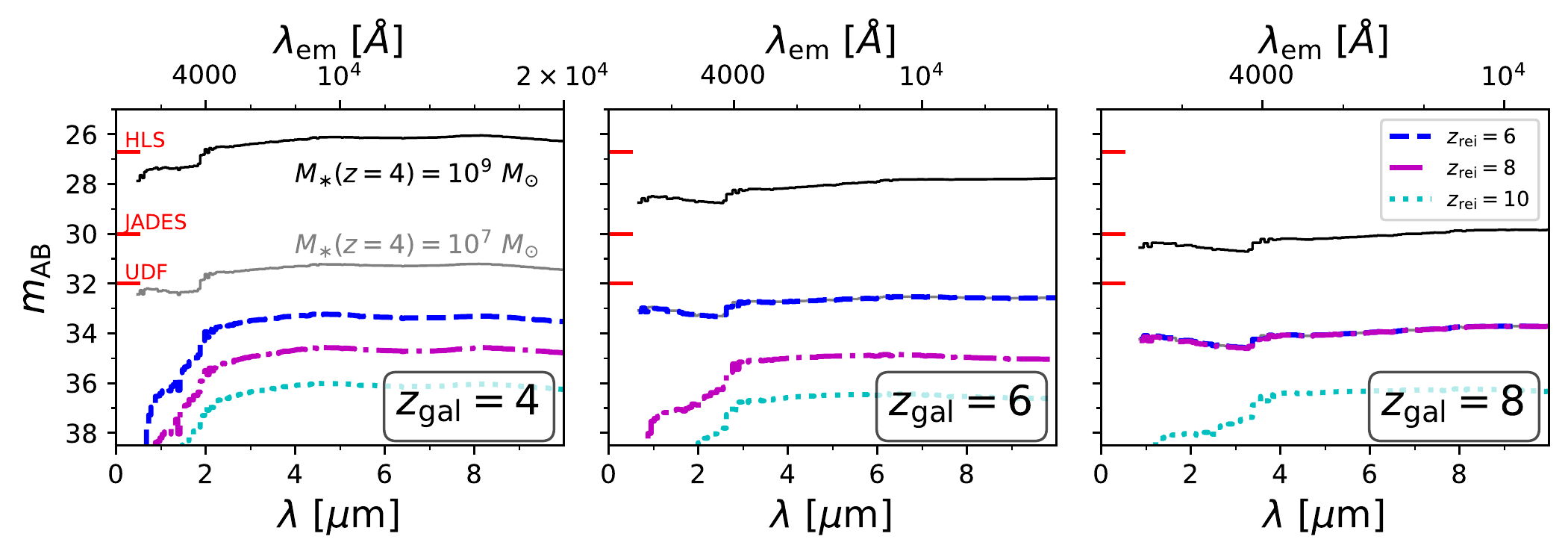}
\caption{{\bf Quenched galaxies have apparent magnitudes $m_{\mathrm{AB}} \gtrsim 33$, beyond the limits of any planned survey at $z \gtrsim 4$.} From left to right, we show the spectra of modeled galaxies (stellar continuum only) `observed' at $z_{\mathrm{gal}} = 4,6,$ and 8 respectively, including an unquenched galaxy with $M_{\ast} = 10^9 \Msun$ (solid black; $M_h \simeq 5 \times 10^{10} \ \Msun$), as well as a series of spectra for $M_{\ast} = 10^7 \ \Msun$ galaxies ($M_h \simeq 4 \times 10^{9} \ \Msun$). For the low-mass galaxy case, we first show a no-quenching scenario (gray), then instantaneous quenching at $\zrei = 6,8,$ and 10 (blue, magenta, and cyan, respectively). Note the strong decline in rest-UV emission ($\lambdaObs \lesssim 2-3 \microns$ depending on $z$) for faint galaxies relative to the rest optical. Nominal magnitude limits for the HLS, JADES, and UDF indicated along the vertical axis, apply at $\lambda \simeq 1 -2 \ \microns$.}
\label{fig:aging_seds}
\end{center}
\end{figure*}

With a set of halo merger trees covering all $z \gtrsim 5$, we synthesize the spectrum of each model galaxy, indexed with $i$, by integrating over its star formation history (SFH), i.e.,
\begin{equation}
    L_{\lambda,i} (\zobs) = \int_{\zobs}^{\zform} \dot{M}_{\ast,i}(\zpr) l_{\lambda}(\Delta t^{\prime}) \bigg|\frac{dt}{dz} \bigg| dz \label{eq:synth}
\end{equation}
where $l_{\lambda}(\Delta t^{\prime})$ is the luminosity of a simple stellar population of age $\Delta t = t(\zobs) - t(\zpr)$ and $\dot{M}_{\ast,i}(\zpr)$ is the SFR of galaxy $i$ at redshift $\zpr$. This luminosity is then reddened by an optical depth $\tau_{\lambda,i}$, yielding the observed luminosity
\begin{equation}
    L^{\prime}_{\lambda,i} = L_{\lambda,i} (\zobs) e^{-\tau_{\lambda,i}} \label{eq:reddening}
\end{equation}
where the optical depth is computed as
\begin{equation}
    \tau_{\lambda} = \kappa_{\lambda} N_d = \kappa_{\lambda} \frac{3 M_d}{4 \pi R_d^2} \label{eq:tau_d} .
\end{equation}
Here, $\kappa$ is the dust opacity, which we assume to be $\propto \lambda^{-1}$, comparable to an SMC-like dust law in the rest-UV \citep{Weingartner2001}. The dust column density, $N_d$, can be recast in terms of a dust scale length and dust mass; we assume that the dust mass is equal to a constant fraction of each galaxy's metal mass, $M_d = 0.4 M_Z$, and parameterize the dust scale length $R_d$ with a double power-law whose parameters are determined via empirical calibration as described above \citep[see Table 1 of][]{Mirocha2020a}. At face value, this model represents a spherically-symmeteric dust screen surrounding a point source of radiation, in which the scale length sets both the mass density and spatial extent of the dust. We do not adhere strongly to this interpretation, and instead think of this as a simple way to model an \textit{effective} dust optical depth or dust column density.

We adopt the \textsc{bpass} version 1.0 \citep{Eldridge2009} single-star models throughout when modeling $l_{\lambda}(\Delta t^{\prime})$ with an intermediate stellar metallicity of $Z=Z_{\odot}/5 = 0.004$. Nebular continuum is included following a relatively standard procedure \citep{Ferland1980,OF2006} \citep[see][for the \textsc{ares} implementation]{Sun2021}, and nebular line emission is neglected in this work for simplicity\footnote{While it is relatively straightforward to include $\Lya$ emission, the many rest-optical lines in the band of interest require more detailed modeling via, e.g., \textsc{cloudy} \citep{Ferland1998,Ferland2013}. We defer such a treatment to future work, but note here that the inclusion of nebular lines should only increase the contrast between quenched and unquenched galaxies, i.e., our current treatment is conservative in this regard.}. Finally, we apply the \citet{Madau1995} model for absorption by intervening clouds in the low-$z$ IGM.

% JTM: talk about Ly-a here?
% Talk about application of Madau model for z>6 sources?

Examples of model galaxy spectra are shown in Figure \ref{fig:aging_seds}. We focus on two illustrative cases: (i) a galaxy unaffected by feedback (solid black), with $M_{\ast} = 10^9 \ \Msun$ at $z=4$ ($M_h \simeq 5 \times 10^{10} \ \Msun$), and (ii) a less massive galaxy with $M_{\ast} = 10^7 \Msun$ at $z=4$, which occupies a $M_h \lesssim 10^9 \ \Msun$ halo at all $z \gtrsim 4$ (gray). The brighter galaxy is detectable at $\zgal=4$ for the \textit{Roman} high-latitude survey (HLS), but is only detectable at higher redshifts with JWST, either in the medium-deep JADES \citep{Williams2018} or an ultra-deep field (UDF) with nominal limiting magnitude $\mAB = 32$. The fainter galaxy, while just bright enough to be detected at $z=4$ in a JWST UDF, is too faint to be detected if its star formation is quenched instantaneously (i.e., its SFR set to zero) at $\zrei=6,8,$ or 10 (dashed, dashed-dotted, and dotted curves, respectively). The rapid decline of the rest-ultraviolet continuum in the quenched galaxy drives a strong $4000\angstrom$ break (see top axis), which is visible at $\sim 2-3 \ \mu\rm{m}$ in the observer frame. Note also that at $\lambdaObs \sim 1 \mu\rm{m}$ the quenched galaxy is even fainter, $\mAB \lesssim 36$, justifying the neglect of aging effects in earlier studies focused on the shorter wavelength end of the near-IR bandpass \citep{Fernandez2012}.

\begin{figure*}
\begin{center}
\includegraphics[width=0.98\textwidth]{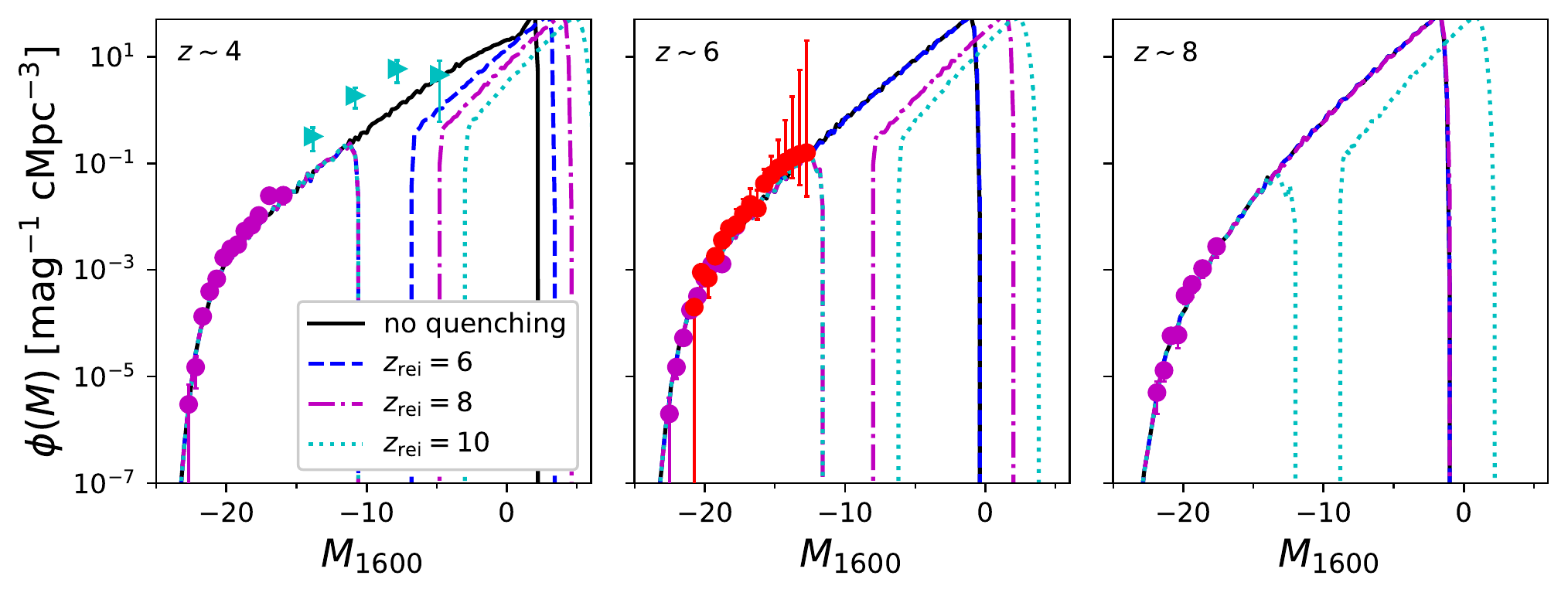}
\caption{{\bf The effects of reionization are likely beyond the reach of high-$z$ UVLFs,} as shown here using a toy instantaneous quenching model. Linestyles indicate the quenching model, from no quenching (solid), to quenching that occurs instantaneously at $\zrei=6,8$, and 10 (dashed, dotted, dash-dotted) in all halos below $10^9 \ \Msun$. Data shown are from \citet{Bouwens2015} (magenta) and \citet{Bouwens2017} (red), the latter showing the power of lensed fields. In addition, at $z=4$ we show the UVLF predicted using reconstructed SFHs of dwarf galaxies in the local group \citep[cyan triangles;][]{Weisz2014FossilRecord}.}
\label{fig:aging_uvlfs}
\end{center}
\end{figure*}

In Fig. \ref{fig:aging_uvlfs}, we show an illustrative example of reionization feedback's impact on galaxy UVLFs, in which the position dependence of the feedback is neglected. Instead, we quench all galaxies with $M_h < 10^9 \ \Msun$ instantaneously for reionization redshifts $\zrei=6,8$, and $10$. As expected from previous work \citep[e.g.,][]{Yue2016a}, a turn-over appears in the UVLF at $-12 \lesssim \MUV \lesssim -10$, with the quenched population of galaxies drifting rightward with time as the stellar continuum fades and reddens. Earlier quenching widens the gap between the quenched and unquenched populations as time goes on. This toy model is compatible with constraints from lensed fields which reach $\MUV \simeq -12$ \citep{Bouwens2017}, but are as yet inconclusive with regards to any turn-over in the UVLF \citep{Atek2018}. Local group constraints from \citet{Weisz2014FossilRecord} (cyan points at $z\sim 4$) provide tantalizing evidence of a UVLF that extends to $\MUV \simeq -5$, in contrast to this simple quenching scenario.

%%
% EoR quenching
\subsection{Reionization \& Quenching} \label{sec:quenching}
In all that follows, we employ a more physically-motivated model for reionization quenching than that used for illustrative purposes in Figures \ref{fig:aging_seds} and \ref{fig:aging_uvlfs}. To model reionization, we use a simple semi-numerical algorithm motivated by the excursion set approach \citep{Furlanetto2004}, of which there are many implementations available in the literature. In short, we first compute the hydrogen-ionizing luminosity density in coarse voxels ($2 \ h^{-1} \ \mathrm{cMpc}$ on a side), including photons generated by all halos resolved by our $N$-body simulation that have at least 10 particles and have existed for at least two timesteps (40 Myr). Then, we iteratively smooth the ionizing luminosity box with progressively finer kernels, flagging voxels as ionized when the total number of photons produced exceeds the number of hydrogen atoms, i.e., when
\begin{equation}
  N_{\gamma}^i (z,R) \geq N_{\HI} = \left(\frac{\Omega_{b,0}}{\Omega_{m,0}} \right) \rho_m (1 - Y) m_{\mathrm{H}}^{-1} V_{\mathrm{vox}} \label{eq:xset}
\end{equation}
where $N_{\gamma}^i(z,R)$ is the cumulative number of ionizing photons emitted at $\zpr \geq z$, in the $i^{\mathrm{th}}$ voxel when smoothed on scale $R$, $Y=0.243$ is the primordial helium abundance, $\mH$ is the hydrogen atom mass, $\rho_m$ is the mean matter density, and $V_{\mathrm{pix}}$ is the volume of each voxel. This expreession is analogous to the more familiar condition, $\fcoll \geq \zeta^{-1}$, relating the collapsed fraction to the ionizing efficiency (neglecting recombinations). Our model, using all resolved halos $M_h \gtrsim 10^8 \ \Msun$ and an escape fraction of $\fesc=0.1$ completes reionization at $z \sim 5.5$ (IGM more than 99\% ionized), with a midpoint at $z \sim 7$ \citep[see Fig. 10 and associated text in][for more details]{Mirocha2021b}.

\begin{figure}
\begin{center}
\includegraphics[width=0.49\textwidth]{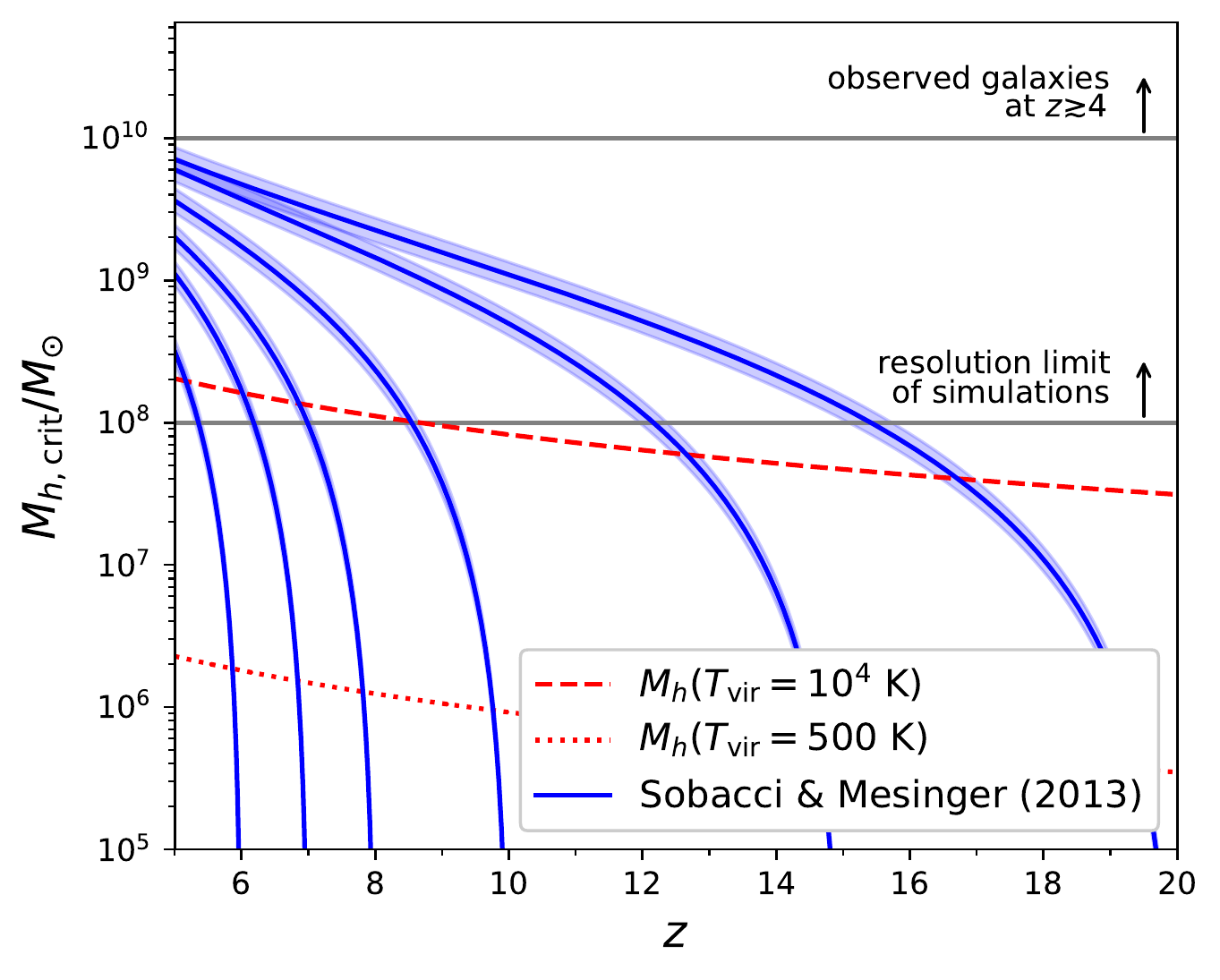}
\caption{{\bf Models employed for critical halo mass below which reionization quenching is effective.} Horizontal gray lines indicate (roughly) the halo masses corresponding to the faintest galaxies observed thus far at $z \gtrsim 4$ (top) and the mass corresponding to a halo containing 10 particles in our N-body simulations (bottom). Blue lines and associated shaded regions are the models of \citet{Sobacchi2013}, adopting a value of $\Gamma=10^{-12 \pm 0.5} \ \mathrm{s}^{-1}$, for a series of different redshifts at which the UV background is assumed to turn on. Red curves correspond to constant halo virial temperatures, including the atomic cooling threshold (dashed), and a $500$ K threshold which corresponds roughly to the smallest molecular cooling halos (dotted).}
\label{fig:mcrit}
\end{center}
\end{figure}

With a reionization realization in hand, we assign each halo a ``reionization redshift,'' $\zrei$, corresponding to the redshift at which its host voxel was flagged as ionized by our filtering procedure. At all subsequent snapshots, $z < \zrei$, we manually set the galaxies' SFR to zero \textit{if} its host halo is less massive than the critical mass. We adopt the fitting formula for the critical suppression mass from \citet{Sobacchi2013},
\begin{equation}
  M_c = M_0 \left(\frac{\Gamma}{10^{-12} \ \rm{s}^{-1}} \right)^a \left(\frac{1+z}{10} \right)^b \left[1 - \left(\frac{1+z}{1+\zrei} \right)^c \right]^d \label{eq:mcrit}
\end{equation}
with $M_0=3 \times 10^9 \ \Msun$, $a=0.17$, $b=-2.1$, $c=2$, and $d=2.5$. Example $M_c(z)$ curves are shown in Fig. \ref{fig:mcrit}. Most importantly in this work, we continue to synthesize the spectrum of all objects starting at $z=10$ down to the last snapshot at $z=5$ following Eq. \ref{eq:synth}. We do so at moderate resolution, $40\angstrom$ in the rest-frame, from 900-7000$\angstrom$. Note that because we set the SFR to zero at all redshifts $z < \zrei$, we are making an optimistic assumption that \textit{all} galaxies with $M_h < M_c$ are quenched immediately, and that they never resume star formation. We will discuss this assumption in \S\ref{sec:conclusions}.

Before moving on, we note that our treatment of reionization feedback is not fully self-consistent. One should in principle perform an iterative procedure to account for the fact that suppressing low-mass galaxy formation will delay reionization. For simplicity, in what follows we adopt the $\zrei$ value for each halo as computed with a fixed suppression mass of $M_h = 10^9 \ \Msun$. The implicit assumption is that the lowest mass halos, which host quenched galaxies, are not the source population driving the bulk of reionization. This is a reasonable assumption for our model, in which the SFE is a steep function of halo mass and $\fesc$ is mass-independent. However, if the SFE is higher in galaxies just beyond current detection limits or $\fesc$ is boosted in low-mass objects, our models would require revision. We defer a detailed exploration of this possibility to future work.

%\begin{figure}
%\begin{center}
%\includegraphics[width=0.49\textwidth]{mock_zrei.pdf}
%\caption{{\bf Reionization redshift averaged along $z$-axis in our fiducial model.} \jtm{Don't think this figure is adding much at this stage but maybe needs to be here?}}
%\label{fig:mock_zrei}
%\end{center}
%\end{figure}

%\begin{figure}
%\begin{center}
%\includegraphics[width=0.49\textwidth]{mock_QHII.pdf}
%\caption{{\bf Mean reionization history in fiducial model.} \jtm{Dunno if we need this figure. Pretty generic.}}
%\label{fig:mock_QHII}
%\end{center}
%\end{figure}

%\begin{figure}
%\begin{center}
%\includegraphics[width=0.49\textwidth]{z_em_v_lam_obs.pdf}
%\caption{{\bf The rest-UV and rest-optical emission of galaxies at $4 \lesssim z_{\rm{em}} \lesssim 10$ corresponds to observed wavelengths of $1 \lesssim \lambda_{\mathrm{obs}} / \mu \mathrm{m} \lesssim 5$.} The dashed curve here highlights the location of the $4000\angstrom$ break, which lies in the $\sim 2-3 \ \mu\rm{m}$ window for galaxies at $4 \lesssim z \lesssim 6$.}
%\label{fig:lam_v_z}
%\end{center}
%\end{figure}

\begin{figure*}
\begin{center}
\includegraphics[width=0.98\textwidth]{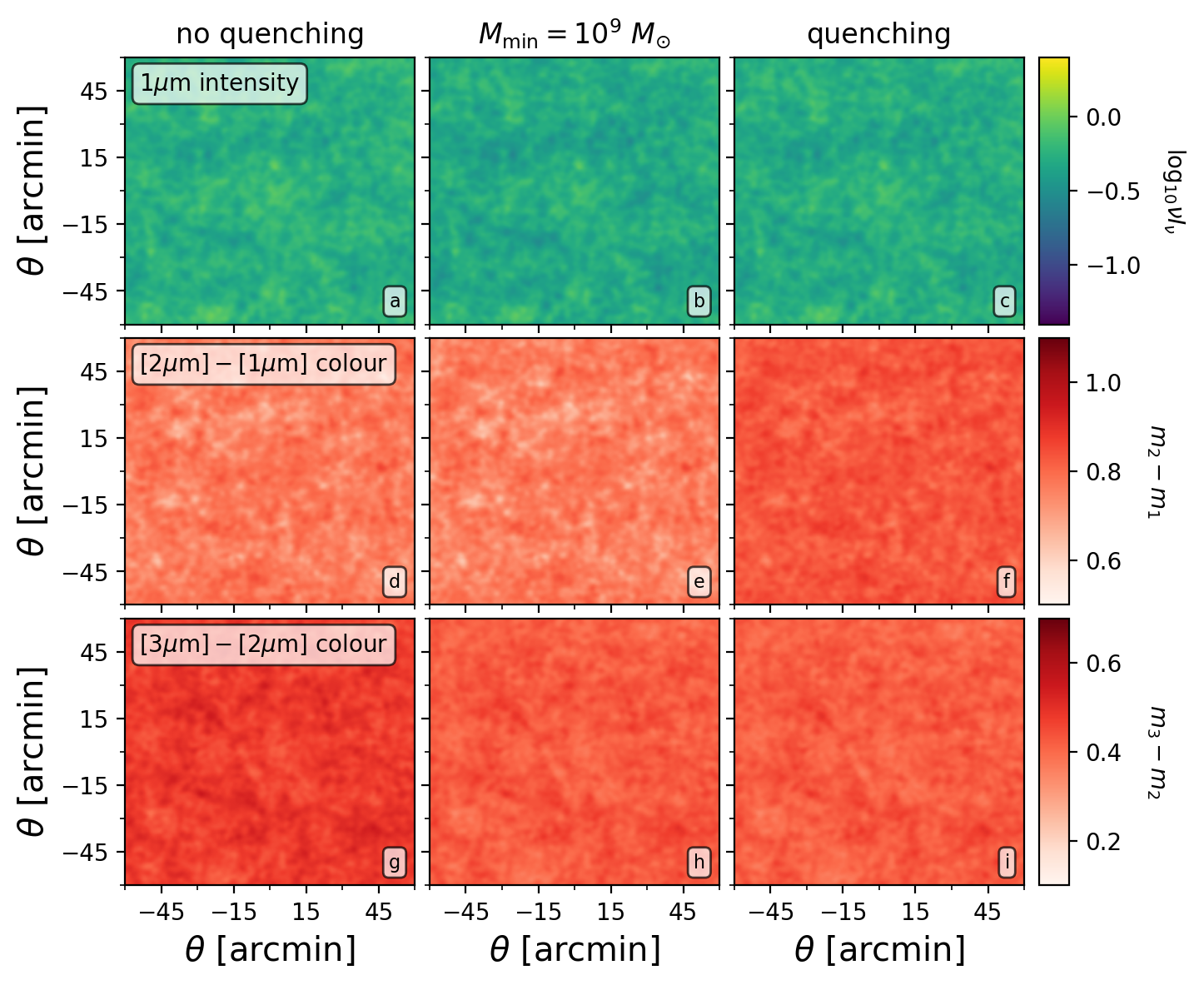}
\caption{{\bf Mock maps of the near-infrared background intensity from $z \gtrsim 5$} at $1 \microns$ (top), with $[2]-[1] \ \microns$ and $[3]-[2] \ \microns$ colours in the middle and bottom rows, respectively. Each column corresponds to a different model, with our quenching-free fiducial model (left), followed by a model with $\Mmin = 10^9 \Msun$ (center), and the physically-motivated reionization quenching model (right). The colours are defined in the conventional way such that more negative values indicate bluer colours. Raw maps are 6 arcsec/pixel -- these have been smoothed with a Gaussian kernel with $\sigma=1$ arcminute width for visualization purposes only.}
\label{fig:mock_nirb}
\end{center}
\end{figure*}

%%
% NIRB
\subsection{Near-Infrared Background Synthesis} \label{sec:nirb}
With a catalog of model galaxies in hand, each with a spectrum synthesized over all $5 \leq z \leq 10$ and rest wavelengths $900 \leq \lambdaEm / \angstrom \leq 7000$, our next step is to construct a light cone at a series of observed wavelengths, $\lambdaObs$, from which we can produce mock images. We perform the following procedure to construct these lightcones:
\begin{enumerate}
	\item Set the angular dimensions of the mock field of view and angular resolution. In all that follows, we adopt 6 arcsecond pixels over a square field of view (FoV) of $(2 \ \rm{deg})^2$. The angular resolution is equal to the SPHEREx pixel scale, though the FoV is much smaller to reduce computational cost.
	\item Set the line-of-sight resolution (4 co-moving Mpc by default).
	\item Populate a light cone with the original halo catalog as viewed from random orientations. We tile the box (again, with random orientations) in order to fill the field of view. For reference, at $z \sim 7$, the simulation box is $\sim 45$ arcminutes in linear dimension, meaning we tile 3 times in each angular dimension, resulting in a FoV slightly larger than $(2 \ \rm{deg})^2$ from which we can extract the patch of interest.
\end{enumerate}
We note that, because of the relatively small 80 $\mathrm{cMpc} \ h^{-1}$ box, we will have repeated structures present in our lightcone. Though this is not ideal, the small size of the box also means evolution effects, i.e., the fact that sources at the front edge and back edge of a given co-eval cube are at slightly different redshifts when populating the lightcone, will be small. In the future, with bigger boxes we plan to generalize our approach to handle such effects, though tiling is likely inescapable for models pushing to $\gtrsim 1 \ \rm{deg}^2$ mocks, especially if faint sources are of interest, as in this work.

We focus our analyses on observed wavelengths $0.8 \leq \lambdaObs / \mu \rm{m} \leq 5$ only, i.e., the same spectral coverage as SPHEREx \citep{Dore2014,Dore2016}. We sample the observed wavelength space with a resolution of $\Delta \lambda = 0.2 \mu \mathrm{m}$, resulting in 22 different lightcones. To make images, we simply sum the background intensity along each sightline, accounting for the cosmological redshift, geometrical dilution of photons, and the IGM opacity using the \citet{Madau1995} model. The resulting images are idealized in that no intervening sources at $z \lesssim 5$ are included, i.e., we effectively assume that foregrounds have already been removed.

In Fig. \ref{fig:mock_nirb}, we show example images produced by the modeling described in this Section. In all that follows, we focus on three models with different treatments of the suppression of the low-mass galaxy population (from left to right in this figure): (i) Our fiducial model with no quenching, (ii) a fixed suppression mass of $\Mmin = 10^{9} \ \Msun$, and (iii) a $\zrei$-dependent suppression mass following the \citet{Sobacchi2013} fitting formula (``quenching''; Eq. \ref{eq:mcrit}).

The top row shows and example intensity maps at $1$ $\microns$, while the bottom two rows show near-infrared colour maps, which compare the intensity at 1 and 2 $\microns$ (middle row), and 2 and 3 $\microns$. By eye, differences in the 1 $\microns$ intensity maps are subtle, though the quenching scenarios do appear to exhibit ``patchier'' features, as expected. The main results of this paper are more clear in the bottom two rows. At the short wavelength, $1-2 \ \microns$  edge of the band explored here, the quenching scenario (panel f) looks qualitatively different than the other two panels (panels d and e). Most notably, the average colour in the quenching image is much redder (less negative) than the others, which makes sense: the stellar populations in the other models are all young. The pure mass quenching model suppresses low-mass galaxy formation, but suppressed halos \textit{never} form stars, so there is no new population of aging objects. In contrast, once reionization quenching is turned on, a new population of objects with intrinsically redder colours is born.

Finally, in the redder wavelength channels (bottom row), it is the ``no quenching'' scenario that stands out as different from the others (panel g). Because the stellar aging effects here are less severe, as we move into the rest optical emission of high-$z$ sources, the reionization quenching scenario will more closely resemble a model in which the population of emitting objects has simply been reduced. Indeed, the quenching model looks more similar here to the $\Mmin = 10^9 \ \Msun$ model.

We examine these maps quantitatively in the next section.

%%%
%% Results
%%%
\section{Results} \label{sec:results}
\begin{figure*}
\begin{center}
\includegraphics[width=0.98\textwidth]{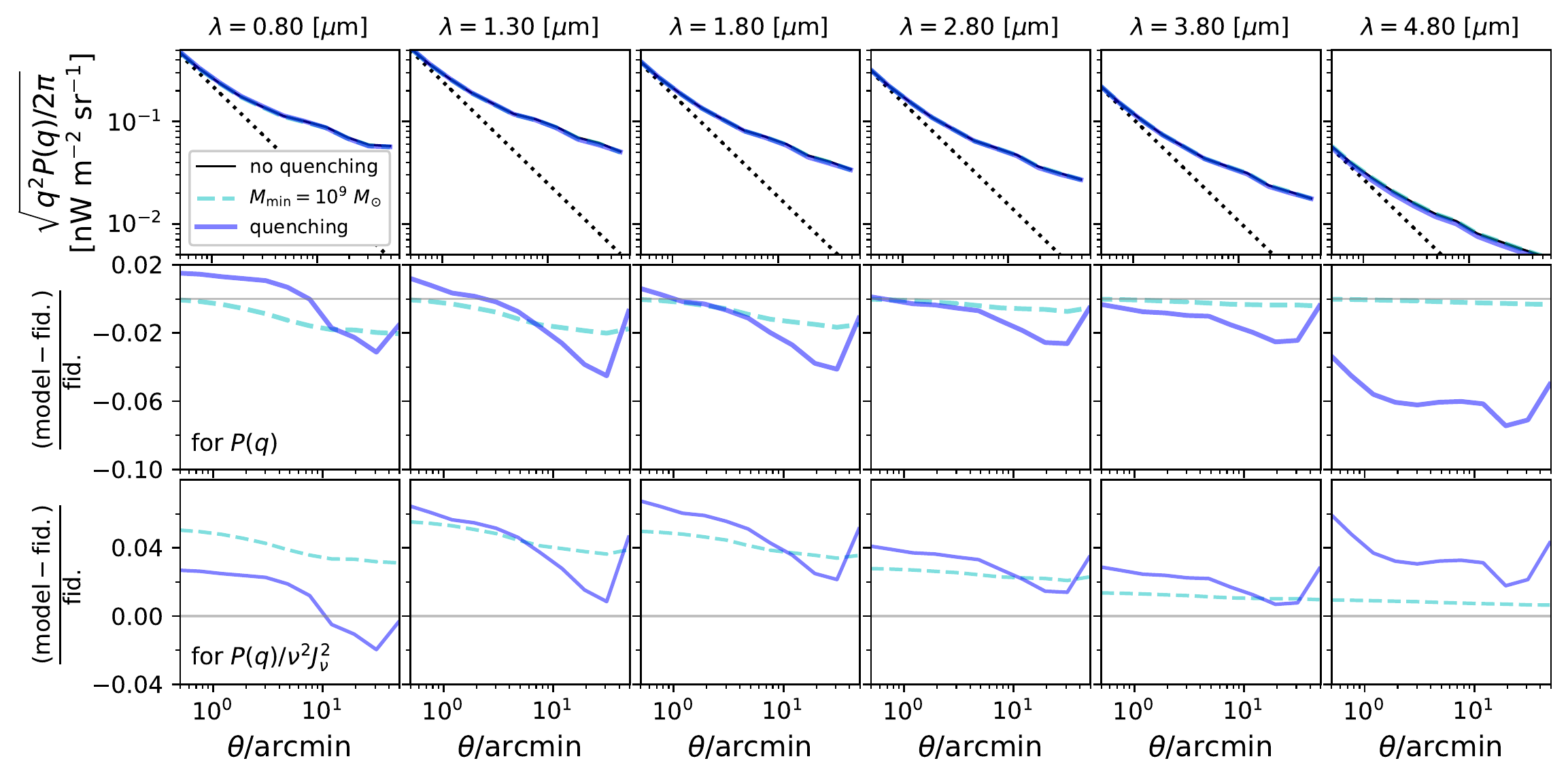}
\caption{{\bf Power spectra of the near-infrared background from 0.8 to 4.8 $\microns$ as a function of angular scale $\theta=2\pi/q$.} Power spectra themselves are shown in order of increasing observed wavelength from left to right in the top row, while the relative differences between each model and the fiducial model, defined as $(\rm{model} - \rm{fiducial}) / \rm{fiducial}$, are shown in the bottom two rows: first, using $P(q)$ itself (middle), and second, focusing on the fractional fluctuations, $P(q)/\nu^2J_{\nu}^2$ (bottom). The shot noise level of the fiducial model is shown in each panel of the top row to guide the eye (dotted black lines). Differences between the fiducial model and quenching models are largest on $10 \lesssim \theta / \rm{arcmin} \lesssim 40$ scales, and while the mass quenching model tends toward the fiducial model at long wavelengths, $\sim 2-5$\%-level differences persist out to $4.8 \ \mu\rm{m}$ for reionization quenching. }
\label{fig:nirb_ps}
\end{center}
\end{figure*}

To examine the signatures of feedback more closely, we proceed to a statistical analysis of the spatial fluctuations in each map, beginning with the power spectrum of individual maps in Figure \ref{fig:nirb_ps}. Consistent with previous studies \citep{Cooray2012,Fernandez2012,Helgason2015,Sun2021}, we predict fluctuations of order $\sim 10^{-1} \ \rm{nW} \ \rm{m}^{-2} \ \rm{sr}^{-1}$ on arcminute scales and short wavelengths $\simeq 1-2 \ \microns$, falling to $\simeq 10^{-2}$ as $\lambda \rightarrow 5 \ \microns$. Current measurements from, e.g., \textsc{AKARI} \citep{Matsumoto2011}, \textsc{CIBER} \citep{Bock2013,Zemcov2014}, HST/NICMOS \citep{Thompson2007,MitchellWynne2015}, and \textit{Spitzer} IRAC \citep{Kashlinsky2012} indicate fluctuations at the $\simeq 1-10 \ \rm{nW} \ \rm{m}^{-2} \ \rm{sr}^{-1}$ level, suggesting that EoR galaxies contribute $\sim 1-10$\% of the NIRB fluctuations.

As expected, each curve in Fig. \ref{fig:nirb_ps} exhibits shot noise on scales $\lesssim 1$ arcminute, with the clustering term becoming important on larger scales (top row). By eye, the two quenching models explored in this work differ little from our fiducial scenario. However, the relative difference between quenching scenarios and the fiducial model (bottom two rows) shows that quenching models differ as a function of angular scale by up to $\sim 2-5$\% at any given wavelength, with the largest differences occurring on large $\theta \simeq 20$ arcminute scales. Note that there is a net suppression of power visible on large scales in the relative difference panels (middle row), which is driven by the fact that the mean intensity is lower in each quenching scenario. This effect is illustrated in the bottom row of Fig. \ref{fig:nirb_ps}, where we compare power spectra that have been normalized by their mean intensity squared, $P(k)/\nu^2 J_{\nu}^2$. Here, one sees that indeed fluctuations are boosted in both quenching scenarios, owing to the higher mean bias of sources.

%The most important result shown in Fig. \ref{fig:nirb_ps} is that, while the fixed $\Mmin = 10^9 \ \Msun$ scenario tends toward the fiducial model as wavelength increases (left to right), the reionization quenching scenario is consistently different at the $\sim 2-5$\% level, with a scale-dependence that differs from the fixed $\Mmin$ model as well.

\begin{figure*}
\begin{center}
\includegraphics[width=0.98\textwidth]{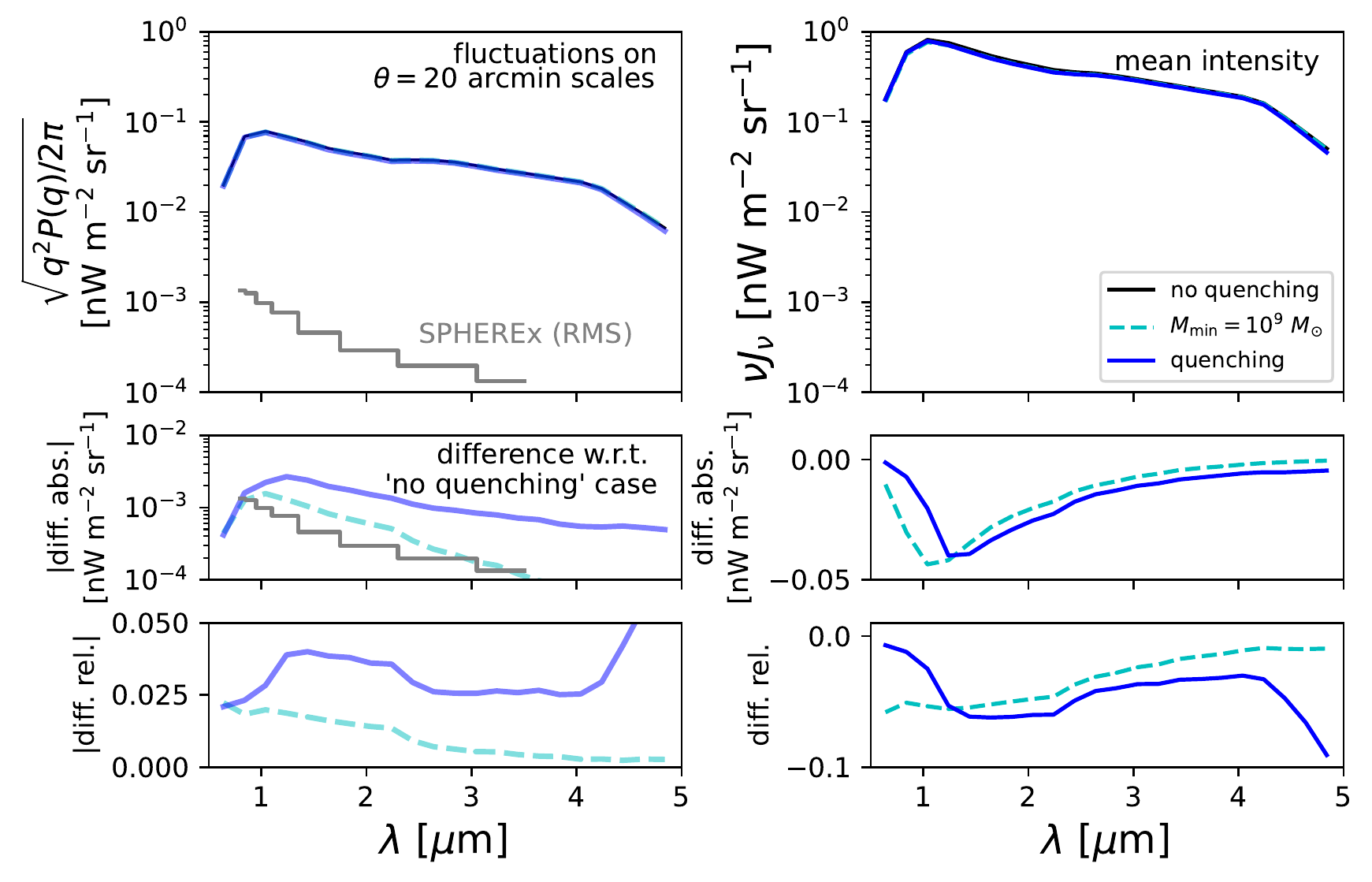}
\caption{{\bf Signatures of feedback in the NIRB power spectrum (left column) and mean (right column) are comparable to the nominal SPHEREx sensitivity.} Top row shows model predictions, while the absolute and relative differences between the fiducial model and all quenching models are shown in the middle and bottom rows, respectively. The SPHEREx deep-field sensitivity (RMS on $5 \leq \theta /\mathrm{arcmin} \leq 22$ scales) is shown in gray for reference \citep[from][]{Feng2019}.}
\label{fig:nirb_colour}
\end{center}
\end{figure*}

To build intuition for these results in more detail, we highlight five key trends apparent in Fig. \ref{fig:nirb_ps}:
\begin{enumerate}
    \item The amplitude of NIRB fluctuations are strongest at the shortest wavelengths, regardless of the model (top row; scanning left to right). This reflects the blue intrinsic colours of high-$z$ galaxies and the rising cosmic SFRD with cosmic time, which gives more weight to the rest-UV emission from sources at the low edge of the redshift interval we study.
    \item The reionization quenching model drives a sharper decline in large-scale power, $\theta \gtrsim 10 \ \rm{arcminutes}$ ($\sim 25 \ h^{-1} \rm{cMpc}$ during reionization), than the fiducial or fixed $\Mmin=10^9 \Msun$ models. This is the imprint of the bubble scale -- sources effectively become less biased on these scales since galaxies outside bubbles are unaffected by reionization. As a result, the fluctuations on the largest scales are always weaker in the reionziation quenching model than in the mass quenching model.
    \item The fixed $\Mmin=10^9 \Msun$ model produces stronger fluctuations than the quenching model at short wavelengths, $\lambda \simeq 0.8 \ \microns$. This is because halos with $M_h < \Mmin$ never exist in the pure mass quenching model, which maximizes the source bias. In the reionization quenching model, many galaxies will form a few generations of stars before being quenched, and so the effective bias of sources will be lower, as will the shot noise.
    \item The fixed $\Mmin=10^9 \Msun$ model tends toward the fiducial model as $\lambda \rightarrow 5 \ \microns$ (middle row; scanning left to right), even after accounting for the overall decline in the mean intensity with $\lambda$ (bottom row). At these long wavelengths, one can think of the NIRB as a tracer of stellar mass more than SFR. Given that the peak in the stellar mass -- halo mass relation occurs at higher $M_h$ than the peak in the SFR-$M_h$ relation, we expect the redder channels to be less sensitive to $\Mmin$.
    \item The reionization quenching signature persists at all wavelengths. We expect a broad-band signature given that the Balmer break is itself a broad-band feature, and the extended time interval over which quenching occurs acts to further broaden the wavelength channels sensitive to quenched galaxy populations. However, we caution the reader that the results in $\lambda \gtrsim 4 \ \microns$ regime may be affected by our relatively small simulation volume. As shown in Fig. \ref{fig:aging_seds}, these channels probe the rest-ultraviolet continuum of $z > 8$ sources. Since the bulk of reionization occurs at $z < 8$ in our model, any galaxies quenched at such high redshifts will be in the rarest overdensities, which are sampled poorest by our $(80 \ \rm{cMpc} \ h^{-1})^3$ box.
\end{enumerate}
In Figure \ref{fig:nirb_colour}, we move on to the colour of the NIRB, both in the power spectrum (on $20$ arcminute scales; top left) and the mean background (top right). Here, we show both the absolute (middle) and relative (bottom) differences between quenching models and the fiducial case without quenching. For the absolute difference, we compare to the nominal sensitivity in the SPHEREx deep field \citep[from][]{Feng2019}, and find that reionization feedback effects are in principle detectable. In the bottom panel, we see once again that reionization feedback induces $\sim 2 - 5$ percent level changes in the power spectrum, while in the mean background (bottom right) the effect is $\sim 2-3$x larger. Once again, the reionization feedback case differs consistently throughout the near-infrared band, while the fixed $\Mmin$ model approaches the fiducial model as $\lambda$ increases.

%%
%% Power spectra
Given that reionization feedback is expected to differentially affect the colours of galaxies above and below the quenching threshold, we now turn our attention to fluctuations in the \textit{ratios} of NIRB intensity maps, which are equivalent to the power spectra of colour maps, like those shown in Fig. \ref{fig:mock_nirb}. We define the power spectrum of map ratios as $P_{1/2}$, i.e., the map at wavelength $\lambda_1$ is in the numerator of the ratio, while the map at $\lambda_2$ is in the denominator, and refer to $P_{1/2}$ as the flux ratio power spectrum (FRPS) from here onward. Note that for these power spectra, the scaled power spectrum $q^2 P_{1/2}$ is dimensionless.

\begin{figure*}
\begin{center}
\includegraphics[width=0.98\textwidth]{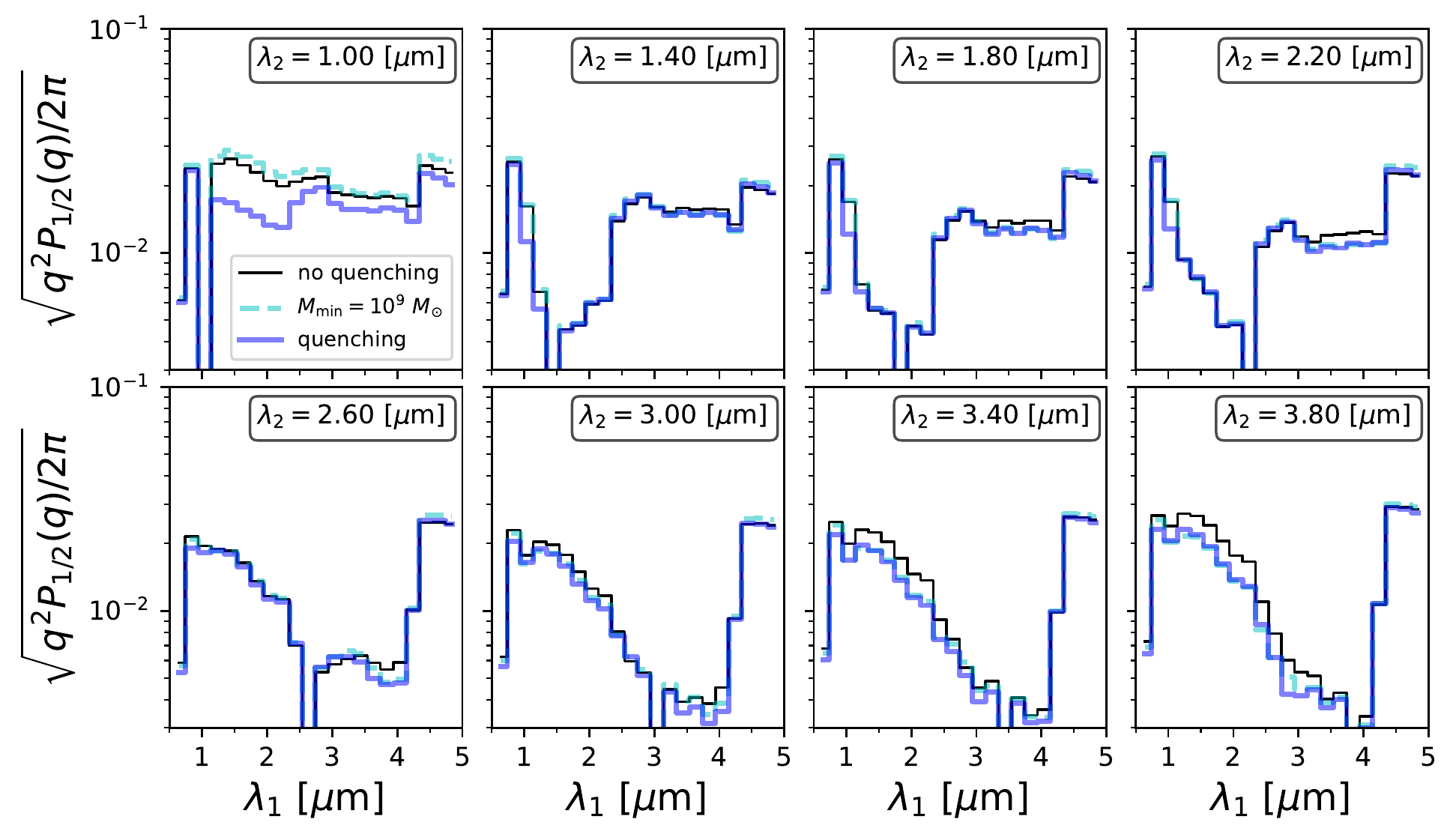}
\caption{{\bf Power spectra of near-infrared background intensity \textit{ratio} maps on 20 arcminute scales.} From top-left to bottom-right, we show the FRPS, $P_{1/2}$, as a function of $\lambda_1$ for several choices of $\lambda_2$ from $1 \leq \lambda_2/\microns \leq 3.8$. Our fiducial model without feedback is shown in black, while constant $\Mmin$ and reionization quenching models are shown in cyan and blue, respectively (see legend in upper-left panel).}
\label{fig:nirb_frps_v_wave}
\end{center}
\end{figure*}

Figure \ref{fig:nirb_frps_v_wave} shows the FRPS on 20 arcminute scales as a function of $\lambda_1$ for a series of different $1 \leq \lambda_2 / \microns \leq 3.8$. Here, unlike the power spectrum itself, differentiation among models is visible by eye. We might expect the strongest signature to occur when taking the ratio of one map that largely probes the rest-optical emission and another map that largely probes the rest-ultraviolet; both wavelength regimes will be suppressed in quenched galaxies, though the rest-ultraviolet will be suppressed more significantly. Rest wavelengths $\lambdaEm \leq 4000 \angstrom$, where quenching effects are strongest, map to observed wavelengths
\begin{equation}
	\lambdaObs \lesssim 2.8 \ \microns \ \left(\frac{\lambdaEm}{4000 \angstrom} \right) \left(\frac{1+z}{7}\right) .
\end{equation}
This heuristic argument is not obviously born out in Fig. \ref{fig:nirb_frps_v_wave} -- the quenching model differs most from the others when $\lambda_1 \lesssim 2.5 \microns$, and $\lambda_2 = 1 \microns$, both of which will be dominated by the rest-ultraviolet emission of high-$z$ galaxies. We defer a more detailed exploration of these trends to future work.
%As a result, flux ratios with $\lambda_2 \lesssim 2.8 \microns$ should be boosted, especially when $\lambda_1 \gtrsim 2.4 \microns$, since these ratios are taking channels largely unaffected by feedback and dividing by channels for which the contribution from faint galaxies is suppressed, thus driving up the flux ratio $f_1 / f_2$ in pixels where feedback has occurred. \jtm{Is this explaining the figure at all?}

Finally, it is perhaps not surprising that the FRPS is more sensitive to reionization feedback than the PS itself, at least for $\lambda_2 \simeq 1\microns$, given that the PS contains no information about the colour of individual pixels. However, there is a cost: the statistics of \textit{noisy} flux ratio maps have some unfortunate properties. For example, if the distribution of fluxes in a given spectral channel is Gaussian, the ratio of fluxes in two different spectral channels will be a Cauchy distribution, which has much broader tails than a Gaussian. As a result, achieving a high-significance detection of the FRPS may require very low-noise per-pixel measurements of intensity maps. See Appendix \ref{sec:appendix} for more details.

\section{Discussion \& Conclusions} \label{sec:conclusions}
In this work, we synthesized mock maps of the near-infrared background in order to quantify the effect of reionization feedback, and whether it may be detectable with near-future facilities like SPHEREx. We start from an $(80 \ \rm{cMpc} \ h^{-1})^3$ $N$-body simulation that resolves all halos $M_h \gtrsim 10^8 \ \Msun$, and build three different realizations of the galaxy population via semi-analytic modeling: our fiducial case with no quenching of any kind, a case with pure mass quenching in which halos with $M_h < 10^9 \ \Msun$ do not form stars, and a final case in which star formation is quenched in galaxies in accordance with their local reionization redshift, as determined via semi-numerical modeling. In each case, we synthesize the full rest-optical and ultraviolet spectrum of each object over its past star formation history in order to predict the NIRB out to $5 \ \microns$, i.e., we do not just link the NIRB to $\Lya$ emission at high redshifts or the recent star formation history of each object, as is common (and perhaps justified when neglecting feedback) in this context.

The main result of this work is that reionization feedback imparts a small -- but potentially detectable -- signature on the near-infrared background. Both quenching scenarios we explore affect the large-scale power spectrum of the NIRB at the $\sim 2-5$ percent level, just above the SPHEREx sensitivity limit (see Fig. \ref{fig:nirb_colour}). Each case exhibits a distinct dependence on angular scale and colour, with the pure mass-quenching model increasingly similar to the fiducial model at longer wavelengths, while the reionization quenching model remains distinct at the $\sim 2-4$ per-cent level out to $\simeq 5 \microns$. While these effects are scarcely visible by eye in the power spectrum of intensity maps, the power spectra of colour maps do bear visually apparent differences. Extracting this information will be difficult (see Appendix \ref{sec:appendix}), but perhaps possible with next-generation facilities.

Our predictions are in good qualitative agreement with previous predictions for the NIRB signal from reionization-era sources \citep[e.g.,][]{Fernandez2010,Sun2021}. To our knowledge, the NIRB signatures of reionization feedback have previously been explored only by \citet{Fernandez2012}, who focused on the $\sim 1 \ \microns$ anisotropies with and without feedback, with reionization modeled via radiative transfer simulations. The most striking difference in our predictions is the magnitude of the feedback effect; our predictions indicate a much weaker signature. This is most likely caused by the assumption of a constant mass-to-light ratio in \citet{Fernandez2012}, whereas ours is a strong function of halo mass. This choice is motivated by high-$z$ UVLFs, which have dramatically improved our constraints of $z \gtrsim 4$ galaxies in the last decade. However, our understanding of high-$z$ sources is far from complete. For example, departures from a single power-law relationship between halo mass and SFR are not unexpected, even in the atomic cooling halo population. Hints of this are apparent in several semi-empirical models \citep[e.g.,][]{Mason2015,Yue2016a}, and predicted in some models of bursty star formation \citep{Furlanetto2022}. If the efficiency of star formation is indeed boosted in low-mass atomic cooling halos, the signature of quenching could be stronger.

Along the way, we have made a number of simplifying assumptions that may impact our model predictions or interpretation of them. For example, we have optimistically assumed that quenched galaxies immediately cease forming stars, and never again resume star formation over their lifetimes. This is clearly simplistic \citep[see, e.g.,][]{Ricotti2005,Wright2019,Katz2020}; plus, observations of nearby galaxies indicate that star formation persists in some dwarfs even after reionization (see \S\ref{sec:intro}). Any `re-invigoration' of star formation at sufficiently late times has the potential to be spectrally distinct from high-$z$ contributions, given that its contribution to the NIRB will be from actual rest-frame IR photons, not rest-optical or UV photons. However, if reionization feedback only briefly suppresses star formation, the high-$z$ signal may be more complicated than the predictions we provide here.

Two final simplifications are worth pointing out. First, we neglect Pop~III stars, which may provide a source of degeneracy given the huge uncertainties on their properties and formation rates. The `Ly-$\alpha$ bump' driven by their intense UV radiation fields would certainly help isolate the contribution of massive Pop~III stars, but only scenarios with very efficient Pop~III star formation will be detectable with SPHEREx \citep{Sun2021}. Other tracers like \rm{He}\textsc{ii} in principle provide another way to isolate Pop~III contributions \citep[e.g.,][]{Visbal2015b,Parsons2021}. Second, we do not explore any cases in which star formation is suppressed via \textit{internal} feedback, which would be subject to stellar aging effects but presumably have no direct link to reionization. One might niavely expect then to retain something akin to the colour dependence we find here, but a scale dependence more like that from a model with fixed minimum mass. We defer more detailed consideration of these possibilities to future work.

Finally, though the effect of reionization feedback in any individual spectral channel is small, a joint analyses over all channels will in principle provide more constraining power. This of course hinges on the ability to separate the various components of the NIRB. While a challenging problem, there is reason to be optimistic, at least in the extraction of the overall high-$z$ contribution to the NIRB in the presence of intra-halo light and other foregrounds \citep{Feng2019}. As a result, attempts to extract even smaller contributions to the NIRB may be warranted, especially in anticipation of the Cosmic Dawn Intensity Mapper \citep[CDIM;][]{Cooray2019}.

J.M. acknowledges helpful conversations with Anthony Harness, Jason Sun, Adam Lidz, and Jose Luis Bernal, and support through a CITA National Fellowship. A.L. acknowledges support from the New Frontiers in Research Fund Exploration grant program, a Natural Sciences and Engineering Research Council of Canada (NSERC) Discovery Grant and a Discovery Launch Supplement, a Fonds de recherche Nature et echnologies Quebec New Academics grant, the Sloan Research Fellowship, the William Dawson Scholarship at McGill, as well as the Canadian Institute for Advanced Research (CIFAR) Azrieli Global Scholars program. P.L. acknowledges support through a Berkeley Center for Cosmological Physics Fellowship. This work used the Extreme Science and Engineering Discovery Environment (XSEDE), which is supported by National Science Foundation grant number ACI-1548562. Specifically, it used the Bridges-2 system, which is supported by NSF award number ACI-1928147, at the Pittsburgh Supercomputing Center (PSC)  \citep{xsede2014}.

\textit{Software:} numpy \citep{numpy}, scipy \citep{scipy}, matplotlib \citep{matplotlib}, h5py\footnote{\url{http://www.h5py.org/}}, mpi4py \citep{mpi4py1}, and powerbox \citep{powerbox}.

\textit{Data Availability:} The data underlying this article is available upon request.

\bibliography{references}
\bibliographystyle{mn2e_short}

%%%%%%%%%%%%%%%%%%%%%%%%%%%%%%%
%%%%%%%%%%%%%%%%%%%%%%%%%%%%%%%
% APPENDIX
%%%%%%%%%%%%%%%%%%%%%%%%%%%%%%%%
%%%%%%%%%%%%%%%%%%%%%%%%%%%%%%%%
\appendix

\section{Statistical Properties of the Flux Ratio Power Spectrum} \label{sec:appendix}
In this Appendix we examine the statistical properties of the flux ratio power spectrum (FRPS), showing that the non-Gaussian statistics of the FRPS means that high-significance detections may require lower-noise measurements in intensity maps than one might naively expect.

Consider first a toy example where a measurement is made only in a single pixel. (Equivalently, one can imagine that we are neglecting spatial correlations between pixels). Let $f_1$ and $f_2$ be the fluxes at $\lambda_1$ and $\lambda_2$, respectively. Our goal is to write down the probability distribution for $r \equiv f_1 / f_2$. To do so, we assume that the $f_i = b_i \delta + n_i$, where $\delta$ is the overdensity field, $b_i$ is a bias factor that encodes the relevant astrophysics (including the feedback effects that we seek to probe), and $n_i$ is the noise in the $i$th wavelength channel. In other words, for the purposes of building intuition we neglect stochasticity in the astrophysics and assume that our Universe's prescription for galaxy formation and evolution is deterministic. Roughly speaking, one is hoping to measure the quantity $b_1 / b_2$.

Assuming that both the density field and the noise are Gaussian distributed, the joint probability distribution for $f_1$ and $f_2$ is given by
\begin{equation}
\label{eq:pf1f2}
p(f_1, f_2) = \frac{1}{\textrm{det}( 2 \pi \mathbf{C} )^{1/2}} \exp \left[ -\frac{1}{2} ( f_1, f_2) \mathbf{C}^{-1}
\begin{pmatrix}
f_1\\
f_2
\end{pmatrix}  \right],
\end{equation}
where we model the covariance matrix as
\begin{equation}
\label{eq:covar}
\mathbf{C}\equiv
\begin{pmatrix}
b_1^2 \sigma_\delta^2 + \sigma^2_1 & \rho b_1 b_2 \sigma_\delta^2 \\
 \rho b_1 b_2 \sigma_\delta^2 & b_2^2 \sigma_\delta^2 + \sigma^2_2
\end{pmatrix},
\end{equation}
where $\sigma_\delta^2$ is the variance of the overdensity field, $\sigma_1$ and $\sigma_2$ are the standard deviations of noise at each of the two wavelengths, and $0\leq\rho\leq1$ is a cross-correlation coefficient between the two bands (reflecting the fact that emission at the two wavelengths may not perfectly trace each other).

The probability distribution $p(r)$ for the ratio $r$ is given by
\begin{equation}
p(r) = \int_{-\infty}^\infty | x| f(rx, x) dx,
\end{equation}
and an explicit evaluation of this expression using Equations \eqref{eq:pf1f2} and \eqref{eq:covar} gives
\begin{equation}
p(r) = \frac{1}{\pi} \frac{\alpha}{\left(r-\beta \right)^2 + \alpha^2},
\end{equation}
i.e., a Cauchy (or Lorentzian) distribution with mean
\begin{equation}
\label{eq:rexpt}
\langle r \rangle = \beta \equiv \frac{\rho b_1 / b_2}{1 + \sigma_2^2 / b_2^2 \sigma_\delta^2}
\end{equation}
and a full-width-at-half-maximum (FWHM) value of
\begin{equation}
\alpha = \left[ \frac{b_1^2 \sigma_\delta^2 + \sigma_1^2}{b_2^2 \sigma_\delta^2 + \sigma_2^2} \left( 1 - \frac{\rho^2 b_1^2 b_2^2 \sigma_\delta^4}{(b_1^2 \sigma_\delta^2 + \sigma_1^2)(b_2^2 \sigma_\delta^2 + \sigma_2^2)}\right)\right]^{\frac{1}{2}}.
\end{equation}

It is instructive to consider the limiting behaviour of this distribution. First, in the high-signal-to-noise regime (where $\sigma_i \lesssim b_i \sigma_\delta$), the mean value of $r$ is given by $\langle r \rangle = \beta \approx \rho b_1 / b_2$.  The result is therefore as expected, in that one recovers $b_1 / b_2$ with $\rho$ as an overall multiplicative factor. In the limit that the two fields trace each other perfectly, $\rho \rightarrow 1$ and $\langle r \rangle = b_1 / b_2$. The FWHM (which we take to be a measure of the error on $r$) in this high-signal-to-noise regime is given by $\alpha \approx (b_1 / b_2) \sqrt{1- \rho^2}$. In the limit that $\rho \rightarrow 0$, we are taking the ratio of two uncorrelated random variables, and with just one pixel in our toy model, the error is on the same order as the signal itself. This is the analog of cosmic variance for our ratio statistic. In the opposite $\rho \rightarrow 1$ limit, the error goes to zero. This is reflective of the fact that if the two bands perfectly trace each other, the stochasticity of the density field cancels out and one is measuring a deterministic quantity (in this case $b_1/b_2$). There is thus no cosmic variance, which is a useful property that has been leveraged in the multi-tracer literature \citep{2009PhRvL.102b1302S,2009JCAP...10..007M,2011MNRAS.416.3009B,2019ApJ...872...82S}.

In the low-signal-to-noise regime, the expectation value of the signal is $\langle r \rangle \approx \rho b_1 b_2 \sigma_\delta^2  / \sigma_2^2$. The numerator of this expression simply reflects the fact that the larger the amplitude of the emission from each wavelength, the more the signal can rise above the noise. It may appear counterintuitive that only $\sigma_2$ (and not $\sigma_1$) appears in the expression. However, if one computes the signal-to-noise ratio, one obtains $\langle r \rangle / \alpha = \rho b_1 b_2 \sigma_\delta^2 / \sigma_1 \sigma_2$, and the noise from both bands enter in a symmetric way.

In the intermediate-signal-to-noise regime, the expectation value of the signal is given by Equation \ref{eq:rexpt}. Importantly, we note that there is a multiplicative noise bias whenever the noise term in the denominator is non-negligible. This bias occurs because in forming the quantity $r = f_1 / f_2$, the noise from $\lambda_1$ enters as an additive term in the numerator of $r$. Thus, if the noise in the first band is symmetrically distributed about zero, it does not contribute a noise \emph{bias} (although of course it increases the noise \emph{variance}). On the other hand, the noise in the second band enters additively in the denominator. This means that even if the noise in this band is symmetric, its effect on $r$ is not, resulting in a noise bias. In principle, one could predict and undo this bias; in practice, given that clean predictions for this bias are only as good as our model, we instead recommend running Monte Carlo simulations.% as we have done in this paper.

A crucial realization in the interpretation of observational constraints on $r$ is the non-Gaussian nature of Equation \eqref{eq:rexpt}. In particular, the Cauchy/Lorentzian distribution has heavy tails, which makes it difficult to distinguish between models at very high significance. Consider, for example, the probability enclosed between $\pm \Delta r$ of $\langle r \rangle$. This is given by
\begin{equation}
P(\beta- \Delta r \leq r \leq \beta + \Delta r) = \frac{2}{\pi} \textrm{arctan} \left( \frac{\Delta r }{\alpha}\right).
\end{equation}
Inverting this relation, one finds that competing models can be distinguished at $68\%$ confidence if their predictions for $r$ are separated by $\Delta r \approx 1.84 \alpha$. If the probability distribution were Gaussian, this would imply that to distinguish two models at $95\%$ confidence requires the models to be separated in $r$ by twice this amount. In our case, however, the heavy tails of our distribution inflate this requirement to $\Delta r \approx 14 \alpha$! The situation gets even worse at $99.7\%$ confidence, requiring $\Delta r \approx 235 \alpha$. %This trend can be seen in Figure \ref{fig:nirb_sens}, where the $95\%$ credibility regions can be an order of magnitude broader than the $68\%$ credibility regions.

Turning this around, it is clear that in order to rule out (or provide evidence for) models at high levels of statistical significance, it is necessary to reduce the measurement errors on $\Delta r$. One way to do this is to simply reduce the per-pixel measurement noise on one's measurement (i.e., to reduce $\sigma_1$ and $\sigma_2$), which in turn reduces $\alpha$. However, this may not always be practical. An alternative is to average down the errors. There are two approaches to doing this. One is to coherently average images in each band \emph{before} taking their ratio. Of course, averaging randomly chosen images of a random cosmological field will result in all signals being averaged down. Thus, to ensure that signals coherently accumulate, it is necessary to carefully select what is being averaged.%, for instance by performing the stacking analysis described in Section \ref{sec:stacking}.

Both of the aforementioned strategies for reducing the error involve suppressing the noise per pixel. An alternative is to combine data in an incoherent manner. For example, the FRPS can be estimated for a number of independent fields and then combined. Such an operation is incoherent in the sense that the ratio of $f_1$ and $f_2$ is taken \emph{before} data is combined. Crucially, when combining multiple estimates of $r$, one cannot take the mean. This is because the error on an estimate of $\langle r \rangle$ does not average down with an increased number of samples, as the tails of a Cauchy/Lorentzian distribution are sufficiently heavy that frequent outliers prevent convergence. Fortunately, there is a simple remedy to this: instead of the mean, one can simply compute the median, which can be shown to converge to the ensemble average mean as the number of samples approaches infinity.

% averaging properties. NOT for the
% long tails and confidence regions
% noise bias

\end{document}